\documentclass[english]{article}
\usepackage[utf8]{inputenc}
\usepackage[T1]{fontenc}
\usepackage{babel}
\usepackage{amsmath}
\usepackage{graphicx}
\usepackage{fancyhdr}
\usepackage{url}

\begin{document}

\title{Representation of the German transmission grid for Renewable Energy Sources impact analysis}

\author{Mario Mureddu\textsuperscript{1}}

\maketitle
\thispagestyle{fancy}

1. Department of Physics and Earth Sciences, 
Jacobs University Bremen,
Bremen, Germany.  \\{*}corresponding author(s):
Mario Mureddu (m.mureddu@jacobs-university.de)
\begin{abstract}
The increasing  impact of fossil energy generation on the Earth ecological balance is pointing to the need of a transition in power generation technology towards the more clean and sustainable Renewable Energy Sources (RES). This transition is leading to new paradigms and technologies useful for the effective energy transmission and distribution, which take into account the RES stochastic power output. In this scenario, the availability of up to date and reliable datasets regarding  topological and operative parameters of power systems in presence of RES are needed, for both proposing and testing new solutions. In this spirit, I present here a dataset regarding the German 380 KV grid which contains fully DC Power Flow operative states of the grid in the presence of various amounts of RES share, ranging from realistic up to 60\%, which can be used as reference dataset for both steady state and dynamical analysis.

% This is a manuscript template for Data Descriptor submissions to \emph{Scientific
% Data} (http://www.nature.com/scientificdata). The abstract must be
% no longer than 170 words, and should succinctly describe the study,
% the assay(s) performed, the resulting data, and the reuse potential,
% but should not make any claims regarding new scientific findings.
% No references are allowed in this section. 
\end{abstract}

%  "Levelized cost of electricity renewable energy technologies" (PDF). FRAUNHOFER. 2013. Retrieved 6 May 2014.

\section*{Background \& Summary}
% (700 words maximum) An overview of the study design, the assay(s)
% performed, and the created data, including any background information
% needed to put this study in the context of previous work and the literature.
% The section should also briefly outline the broader goals that motivated
% the creation of this dataset and the potential reuse value. We also
% encourage authors to include a figure that provides a schematic overview
% of the study and assay(s) design. This section and the other main
% body sections of the manuscript should include citations to the literature
% as needed \cite{cite1, cite2}. References should be included within the 
% manuscript file itself as our system cannot accept BibTeX bibliography files. 
% Authors who wish to use BibTeX to prepare their references should therefore 
% copy the reference list from the .bbl file that BibTeX generates and paste it 
% into the main manuscript .tex file (and delete the associated 
% \textbackslash{}bibliography and \textbackslash{}bibliographystyle commands).

The recent growth in the installed Renewable Energy Sources (RES) generation is completely changing the paradigms and rules associated with electrical systems management and control. These systems  were installed with a top-down hierarchical approach\cite{GungorBOOK1988}, where the power produced at high voltage levels by big conventional generators was distributed towards the users, located at the low voltage nodes.
With the renewable energy revolution, the power systems are now facing the presence of a large number of small and medium size generators distributed over all the voltage levels, which implies a total redesigning of the network structure. Moreover, it is difficult to drive the system in the correct operative state, due to the intermittent nature of RES power production.

In order to ensure the correct functioning of power systems, new methodologies and concepts have to be found in the fields of electric system dynamics, control and economics\cite{Motter2002,ScalaPhysD2015,Mureddu2015,Dorfler2013,Olmi2014}, very few public power grid datasets exist, limiting the possibility to perform studies on real systems and limiting the reproducibility of the studies which relies on private data. The lack of datasets is due to various factors, including difficulty in gathering data, secrecy of data for national security issues and difficulty in managing the huge amount of data that characterizes the physical and economic state of these systems. 

The known power grid datasets, which can be used for performing dynamical studies on nation-wide systems, can be found on different repositories. The Koblenz network collection project contains a dataset of the US transmission grid at \url{http://konect.uni-koblenz.de/networks/opsahl-powergrid}, which contains the full system topology. MATPOWER\cite{Zimmerman2011} provides various datasets, including different versions of the Polish HV grid and an estimation of the European transmission grid, which description can be found in \cite{Fliscounakis2013}. The most complete dataset of the European grid, including buses, generators and lines description, and a full DC power flow solution is given in \cite{Zhou:2005,Hutcheon2013}; this dataset, which has been used as a starting point for our further improvements, comes in a fully geolocalized way.

These datasets have been used in various studies regarding power grid stability and resilience, mainly based on the topological aspects\cite{Watts1998,Motter2002,ScalaLNCS2013}.

However, the newly proposed methodologies \cite{Pahwa2014,Denholm2007,Nardelli2014} are based on the description of RES as stochastic elements which feed intermittent power into the network, and the realistic positioning and variability of these elements can be very important in describing their impact. Despite the quality and accuracy of most of the existing datasets, they have been produced mainly in the previous decade, when RES generation was still not an issue and include very little (if not none) information about RES generation.  Taking Germany as example, the RES installed capacity increased from 6.3\% in 2000 to 30\% in 2014\cite{Zeitreihen}, and is bound to further increase in the next years. For this reason, it is clear how the previously described datasets cannot be used  to properly describe the effects of RES generation in power systems. 

In view of this situation, I have developed a dataset describing the German transmission grid state in presence of real RES and conventional generation. The dataset is based on the network topology proposed in \cite{Zhou:2005,Hutcheon2013}, enriched with the RES generation present in Germany at the end of 2014, and updated with the 2014 load and RES production data given in \cite{energymap}. The dataset, containing all 380 KV buses, lines and connected generators in Germany, contains the information about the RES generation, aggregated per bus and tech type, the load of each bus and the parameters and dispatch of each  conventional generator, validated by means of DC Optimal Power Flow (DC OPF). Moreover, it contains the estimation of the network status in presence of very high shares of RES up to 60\% of the total load, under certain assumptions described in the methods section. This dataset is connected with the paper \cite{Mureddu2016}, which uses this dataset for analyzing the islanding capabilities of the German transmission grid. We expect that this dataset can be used for performing dynamics, steady state and economic studies on power grid in presence of an high share of RES generation, due to the accurate description of the operative parameters of the system.

\section*{Methods}
% The Methods should include detailed text describing any steps or procedures 
% used in producing the data, including full descriptions of the experimental 
% design, data acquisition assays, and any computational processing (e.g. 
% normalization, image feature extraction). Related methods should be grouped 
% under corresponding subheadings where possible, and methods should be described 
% in enough detail to allow other researchers to interpret and repeat, if required, 
% the full study. Specific data outputs should be explicitly referenced via data 
% citation (see Data Records and Data Citations, below). Authors should cite 
% previous descriptions of the methods under use, but ideally the method 
% descriptions should be complete enough for others to understand and reproduce 
% the methods and processing steps without referring to associated publications. 
% There is no limit to the length of the Methods section.
In order to make the dataset of practical use for simulation purposes, the German power system obtained from the original UCTE grid data\cite{Zhou2005,Hutcheon2013} has been processed and enriched. 
To do so, the data was processed in three main steps, described in detail in the next sections:
\begin{itemize}
\item The data about the German transmission system was isolated from the original dataset.
\item The obtained German system data was merged with the information about RES installed generation in Germany, coming from \cite{energymap}. In this way, it was possible to estimate the amount of RES installed on each transmission node for values of RES penetration $P_\% = \{20, 30, 40, 50 , 60\} \%$.
\item The dispatch of conventional generation was calculated by means of a DC OPF\cite{Stott1974,Stott2009}, for each value of renewable penetration $P_\%$.  
\end{itemize}
Moreover, in order to increase the dataset completeness, the nodes' rotating inertia was estimated. This information is not useful for performing steady state analysis, but can be precious for dynamical studies, mainly based on swing equation formalism.

\subsection*{Isolation of the German system}\label{sec:extractionGermany}
The original dataset consists of the full European grid. However, the data regarding the installed RES generation  is public available for few European nations. Here we use the data about the German system. Since this data is restricted only to the German territory, it was necessary to isolate it from the bigger UCTE dataset, in order to perform the further proposed improvements. 

The isolation of the German grid is not straightforward. In order to properly isolate the system, it is vital to consider the  power flowing in and out the neighboring countries. 
In general, the European system is interconnected at the voltage levels of interest, and it is not possible to model the power flow of interconnected nations independently. This is due to the fact that a change in the state of the network around the borders can cause a redirection of power flows through the border nodes, which can impact on the internal flows.  In order to overcome this issue, it is common practice in this type of datasets to model the foreign exchanges by creating fictitious nodes at the borders which work as power generators or sinks representing the power exchanges with the neighboring nations.
In order to isolate the German network correctly, I have extracted all the nodes belonging to the German power system and all the lines among them, by using the nation identifiers used in the original  UCTE dataset. Also, it is necessary to model the power flow in and from the neighboring countries. In order to do so, the foreign border nodes with bigest flows have been kept in the system, with a load (positive for outflows, negative for inflows) equal to the expected foreign flow, extracted from the original dataset. The used German and foreign nodes are shown in figure \ref{fig:de_grid}.  These nodes model the exchanges with Denmark, Netherlands, France, Switzerland, Austria, Czech Republic and Poland. The values of the border flows have been taken from the original dataset and considered constant during all the further analysis. Moreover, the RES generation facilities present in the original dataset have been not considered, in order to perform the proposed more precise analysis of the RES spatial distribution described in the next sections.

\subsection*{Estimation of RES generation in the system}
Performing precise and reliable studies regarding the impact of RES power supply on power systems requires a realistic spatial and temporal distribution of the existing renewable generation. Although, the currently public available datasets about RES generation are given on aggregate, by measuring the total power production during time in the whole system or in fractions of it.
		Therefore, the effective RES production of each node is difficult to estimate, especially on transmission level. This is due to its strong dependence on the local meteorological situation and non-linear energy conversion factors of the single nodes. So, it is necessary to identify methodologies that allows to estimate the RES power production on each transmission node, by making use of the available information about the system. 

Fortunately, some European states (i.e. Italy (\url{atlasole.gse.it/atlasole/}) and Germany (\url{http://www.energymap.info/}) have made the information about the geographical and technical characteristics of all their installed RES generators publicly available. %Among this datasets, particularly important for our purposes are the information about all RES generators nominal power, power production during last year, technology and geographic position.
Starting from these datasets, it is possible to estimate, by making some assumptions that will be described later, the amount of installed RES $P^{Inst}_{i,t}$ per transmission node $i$, aggregated per technology $t$.
\subsection*{The dataset about RES installed generation} 
The complete information regarding the installed RES generators in Germany can be found on the web site \url{http://www.energymap.info/}. From this web site, it is possible to download a database containing the following information for each RES generator $g_{RES}$:
		\begin{itemize}
			\item the date of installation;
			\item the postcode of its geographical location;
			\item a registration code;
			\item the technology type (Solar, Wind, Biomass, Hydro);
			\item the nominal power (in KW);
			\item the amount of KWh produced in 2013;
			\item the amount of KWh/year produced on average during the time after installation;
			\item its geographic position, given in terms of lon/lat with a precision of 0.01 degrees;
			\item the responsible Transmission System Operators (TSOs) and Distribution System Operators (DSOs).
		\end{itemize}
        
\subsection*{Installed RES estimation} \label{sec:intalledRes}
The RES power is mainly installed on low and medium voltage nodes, but its variability can easily change the system power balance, causing  the loads of the very high voltage buses $i$ to fluctuate. In order to model the impact of such fluctuations in the system, it is necessary to estimate the amount of RES generation supplied by each node $i$.
The correct association of RES generators $g_{RES}$ with high voltage nodes $i$ is not an easy task, because it depends on the full system status and power flows, and it can in principle vary from time to time. Moreover, a complete knowledge of the entire power system regarding all voltage levels is not available, making a fully correct association process impossible. However, power grids are dynamical systems which always change their state during time, and therefore a correct, static description of the flows of power generated by RES cannot exist. Ideally, a study of transmission grids which includes a correct description of RES should include the simulation of all the lowest voltage levels, including a correct description of the installed RES intermittent output, making the analysis almost impossible to perform, even by having access to all the data. 

\begin{figure} 
			\includegraphics[width=\textwidth]{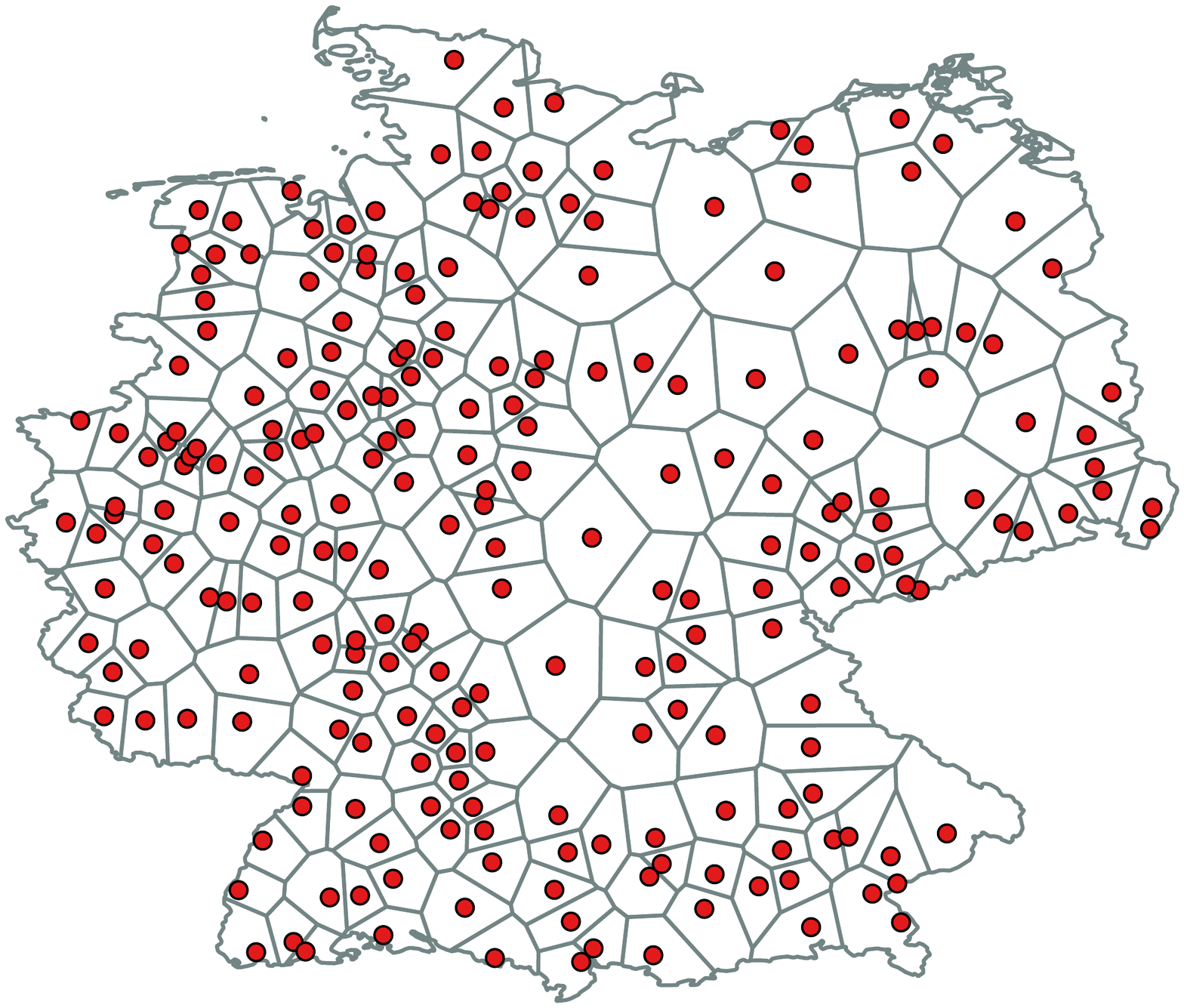} 
			\caption{The Voronoi tassellation of the German territory, with respect to the German 380KV nodes. Each red dot represents a network's node $i$, and its surrounding area is its Voronoi area $A_i$. The RES generators $g_{RES}$, not shown here, were associated with each node $i$ by means of the Voronoi area $A_i$ to which they geographically belong to. The map has been created by using the software QGIS \cite{QGIS}.} \label{fig:voronoi}
		\end{figure}

Consequently, an approximation describing a static association of RES generators with high voltage nodes is necessary for providing a good representation of the system. Given the available data, it is possible to estimate this association on based on the assumption that each RES generator $g_{RES}$ contributes to the power production of the geographically nearest node $i$.  
I made such an assignment by using GIS (Geographical Information Systems) such as PostGIS (\url{http://postgis.net/}) and QGIS \cite{QGIS}. In particular, the area of Germany has been at first divided in 230 zones, each of them corresponding to an internal node, as shown in figure \ref{fig:voronoi}. 

\begin{figure}
			\includegraphics[width=\textwidth]{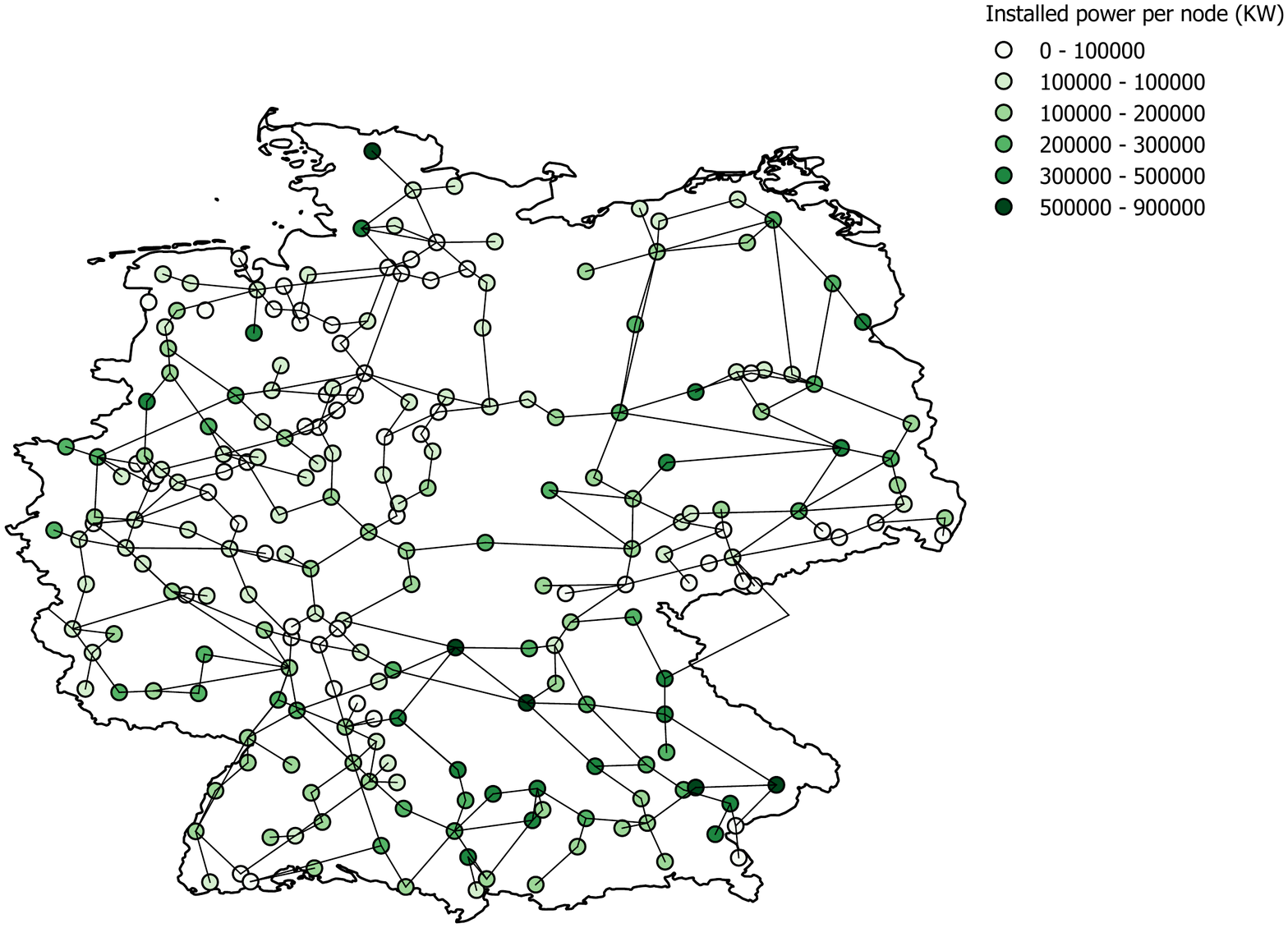} 
			\caption{The map shows the amount of installed PV generation $P*{inst}_{i,PV}$ (in KW)  for each node $i$ of the German transmission network. The association has been made by the nearest node assumption described in section ''Installed RES estimation''. The map has been created by using the software QGIS \cite{QGIS}.} \label{fig:PV}
		\end{figure}

\begin{figure}
			\includegraphics[width=\textwidth]{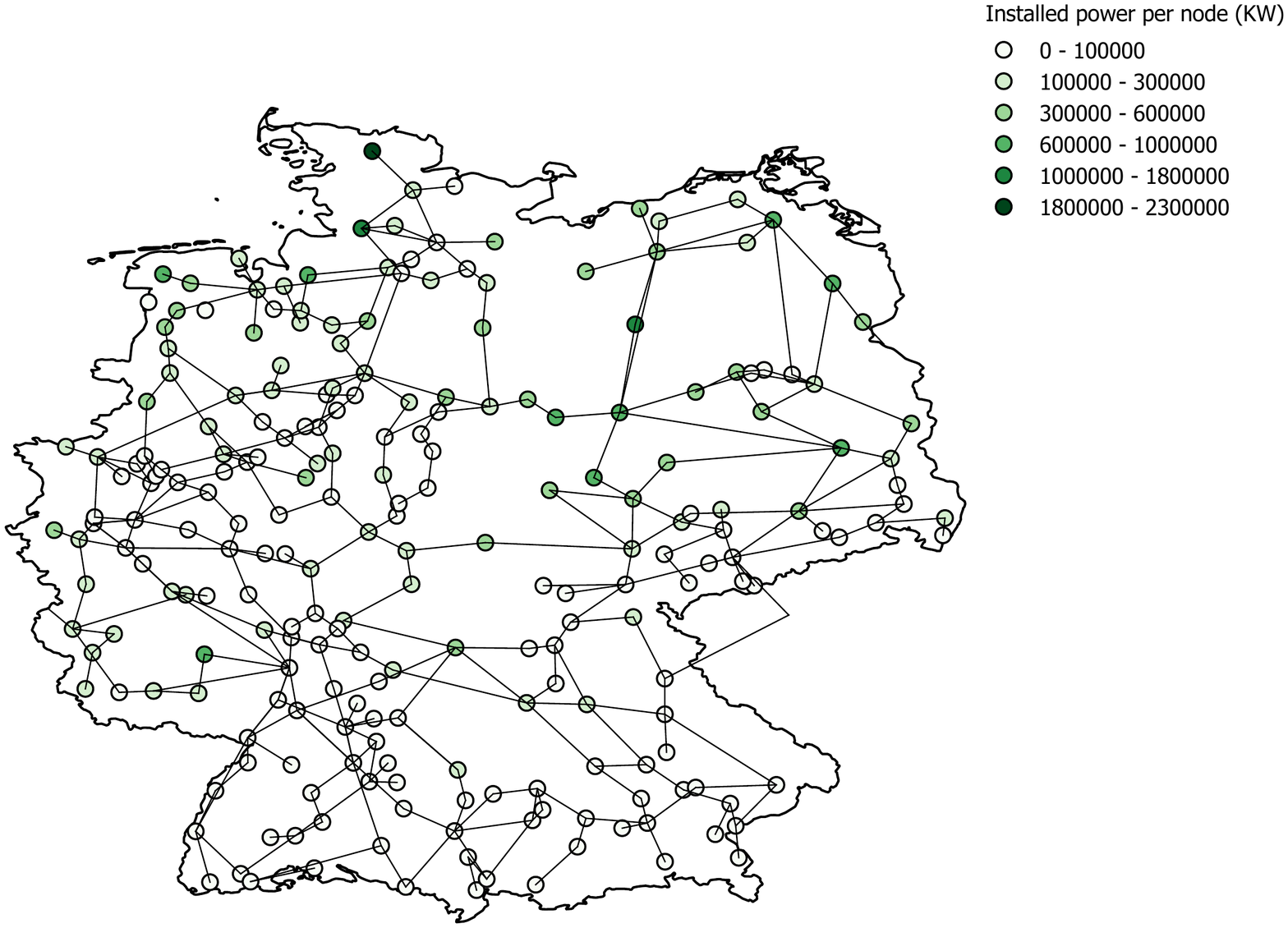} 
			\caption{This maps shows the amount of Wind generation $P*{inst}_{i,Wind}$ which has been assigned to each node $i$, on the base of the nearest node assumption. All powers are expressed in KW.  The map has been created by using the software \cite{QGIS}.} \label{fig:Wind}
		\end{figure}

Such zones have been obtained by means of the Voronoi analysis \cite{Voronoi}. This geographical analysis takes as inputs a set of points ${i}$ distributed in a territory, and produces as output a set of areas ${A_i}$, each of them corresponding to the area closer to the node $i$ than to any other node $j\neq i$. Considering the dataset nodes as the Voronoi points, it was possible to obtain their Voronoi areas $A_i$. 
Once this analysis was performed, the German RES generators $g_{RES}$, which positions are known, were associated with each node by checking which Voronoi area $A_i$ they belong to. In this way, a 1 on 1 association between RES generators $g_{RES}$ and transmission nodes $i$ was obtained. 
After this step, it was possible to sum up all the installed power $P_{i,t}^{Inst}$ in each node $i$, for each type of RES technology $t$. In order to do so, each generator $g_{RES}$ was filtered on the base of its technology and its installed capacity, and all the obtained classes have been summed up for each area $A_i$, representing the node $i$. The results of this analysis are shown in figures \ref{fig:PV} and \ref{fig:Wind}.

		This distribution $P_{i,t}^{Inst}$, representing the amount of installed RES sources for each node $i$ and technology $t$, can be used to realistically describe the wanted amount of RES production through the nodes. In the following section, this procedure will be described in detail.
% As a first approximation, the RES power production can be distributed accordingly to the installed capacities, especially in peak cases. However, for a more precise definition of power outputs during different daytimes, it will be necessary to perform further analysis based on the different technologies technical parameters. In particular, the techs different production/time curves  should be taken into account in the definition of seasonal and daily power output estimation. For such a reason, since the average capacity factor of solar and wind power are comparable, the $P_{i,t}^{Inst}$ have been by a common factor, independently on the tech $t$ \cite{CapFactors}. 

\subsection*{Generation of the power profile}
The original dataset provides a load winter peak solution of the system. However, the dataset includes very limited information regarding RES generation, in the form of few big generators representing RES. Studies regarding dynamics and steady state stability in the presence of RES generation often require as inputs a more detailed description of the system. Therefore, the datasets described in the previous sections were processed in order to obtain a description of the system which includes a realistic representation of RES generation.
This has been done in two steps: estimation of the  RES production and redispatch of the conventional generation. 

The first step makes use of the information regarding the load profile of the daily winter peak for the German system's load $L^{data}$ present in the original dataset. At first, define the share of power $P_\% = {20, 30, 40, 50, 60}$ \% produced by renewable as  $P_\% = \frac{P_{\%}^{tot}}{L^{data}}$, where $P_{\%}^{tot}$ is the power provided in the whole German territory by RES generators of technology $t$. Then, knowing these amounts and the $P_{i,t}^{Inst}$ obtained in the previous section, it is possible to distribute the $P_{\%}^{tot}$ among the German territory. Assuming that all generators of the same RES technology have the same operative parameters (efficiency, wind/sun exposure, and in general capacity factors \cite{CapFactors}) in the entire German territory, it is possible to estimate the RES production per each node $i$, per each technology $t=\{Wind, PV\}$, as defined in Eq. \ref{eq:RES_nodes_gen}. 
\begin{equation} \label{eq:RES_nodes_gen}
P_{i,t}^{\%} = \frac{P_{i,t}^{Inst} * P_{\%}^{tot}}{\sum_i P_{i,t}^{Inst}}
\end{equation}

Again, this assumption is based on the lack of information regarding the capacity factors of all the installed generators and on the extreme complexity associated with a full study of the German weather data during an extended time period.
Once the load consumption and RES generation profiles $L^{norm}$ and $P_{i,t}^{\%}$ are known, it is possible to estimate the power production of conventional generation by means of a DC Optimal Power Flow (OPF) analysis of the system. OPF \cite{Stott1974,Stott2009}, is described in detail in the research paper associated with this dataset\cite{Mureddu2016}. OPF is a standard electrical engineering methodology used for power dispatch (i.e. the decision regarding the patterns of power production in the system).
The inputs for the OPF concerning the conventional generation, such as generators' production prices, min and max power, ramps and voltages, lines' impedances and capacities was taken from the original dataset parameters. The loads' consumption and the RES generation was taken from $L_i$ and $P_i^{RES}$, and it was considered non dispatchable, thus fixed, during the entire OPF process. 

The outcome of each OPF calculation represents the final dataset, with the full distribution of loads $L_i$, the distribution of RES generation $P_{i,t}^{\%}$, and the conventional generation dispatches $P_g^{CONV}$ for each node $i$ and conventional generator $g$. Moreover, the further results of the OPF, such as nodes voltages $V_i$ and phases $\varphi_i$ were inserted in the dataset, since they could be used for further analysis.

		\subsection*{Estimation of rotating inertia}  \label{subsec:inertia}
		The rotating inertia (moment of inertia) of the nodes is difficult to estimate, especially for load nodes. If a conventional generator (CG) is present in the node, the CG turbine rotating inertia could be used as reference value. If this value is not known, as it often happens, it is possible to estimate it from the generator technology and nominal power \cite{Kundur1994}. In particular, as stated in \cite{Kundur1994}, the estimated, average inertia for a conventional generator is expressed in Eq. \ref{eq:inertia1}, where $S_{max}$ is the nominal apparent power, in W of the generator and $k_I^{gen} = 0.5 \frac{s^3}{m}$. Note that if the generator is offline, its inertia contribution must count as zero.
		\begin{equation} \label{eq:inertia1}
		I = k_I^{gen} \cdot S_{max},
		\end{equation}
Moreover, if a load is present on the node, the evaluation of inertia is difficult. The 380KV network is the highest level of the power system hierarchy, and their nodes' load will likely be given by a combination of  both user consumption and generation at lower levels, which in turn could come from RE and conventional sources. For such a reason, a rotating inertia equal to an average motor load has been chosen to represent this nodes. In particular, using the values described in \cite{Kundur1994}, their rotating inertia has been estimated by equation \ref{eq:load_inertia}, where $S_{max}$ is the load load and $k_I^{load} = 0.2 \frac{s^3}{m}$.
		\begin{equation} \label{eq:load_inertia}
		I = k_I^{load} \cdot S_{max}.
		\end{equation}
		In principle, if a node has both a load and a conventional generator, their inertia contribution must be summed or, depending on the needs, it should be split in two different nodes. In general, some datasets tend to represent load nodes with attached generators as mixed nodes. In the particular case of the German grid, however, the original dataset consists on nodes that are either generator or load nodes, and no mixed nodes are present. 
		In general, other effects must be taken into account in the estimation of the inertia. In particular, secondary effects due to electronically controlled devices and control systems and wind turbines should be taken into account. However, a correct estimation of these effects was beyond the scope of this work.

\subsection{Estimation of lines capacity}
The capacity $C_{ij}$ is defined here as the maximum amount of power which can flow through a line before without causing any arm. This quantity, often different from the maximum transmissible power $P_{MAX} = \frac{V^2}{Z}$, is linked to heat production associated to resistive effects in the conductor. In particular, it is considered safe a situation in which the produced heat $H_{prod}$ is less then the dispersed heat $H_{disp}$. Since $H_{disp}$ is dependant from the weather conditions, it is common to define a nominal value of $C_{ij}$, which serves as reference for the daily use of the lines, especially in absence of real time monitoring systems.

The original dataset provided only the information about the $C_{ij}$ of the inter-tie lines. 
Due to the lack of information regarding the other lines capacity $C_{ij}$ in the available datasets, these has been estimated by assuming fixed values for the existing lines, ranging from 800 MW to 4GW, with steps of 400 MW, obtaining the values $C_{allowed} = \{800, 1200, 1600, 2000, 2400, 2800, 3200, 3600, 4000 \}$ MW. These values have been considered a combination of one or more circuits representing the connection. To estimate these values, all lines flows $P_{ij}^{orig}$ calculated in the original dataset have been multiplied by a common factor $c = 1.65$. After this step, the obtained capacities $C_{ij} = c\cdot P_{ij}^{orig}$ have been rounded to the next value in $C_{allowed}$.

\subsection*{Code availability}
The proposed dataset is provided in two different formats: excel spreadsheet and matpower files.

Excel format is one of the most common data formats, highly portable and easily readable by the majority of programs of interest for the field, such as R\cite{RSoftware}, Python (Python Software Foundation, \url{https://www.python.org/}) and Matlab \cite{Zimmerman2011}. Geographical coordinates are saved in lat/lon format. 

Map format is the most common data format for geographical data. It is easily accessible from any GIS software and contains all operative and geographical parameters of the network. The dataset is particularly useful for geographical processing and visualization of the data. A free,  open source program able to open and process this format is QGIS \cite{QGIS}.

% For all studies using custom code in the generation or processing of datasets, 
% a statement must be included here, indicating whether and how the code can be 
% accessed, including any restrictions to access. This section should also include 
% information on the versions of any software used, if relevant, and any specific 
% variables or parameters used to generate, test, or process the current dataset. 

\section*{Data Records}

%Please explain each data record associated with this work, including
%the repository where this information is stored, and an overview of
%the data files and their formats. Each external data record should
%be listed in Data Citation section at the end of this template, and 
%records should be cited throughout the manuscript as, for example 
%(Data Citation 1).

\begin{figure}
			\includegraphics[width=\textwidth]{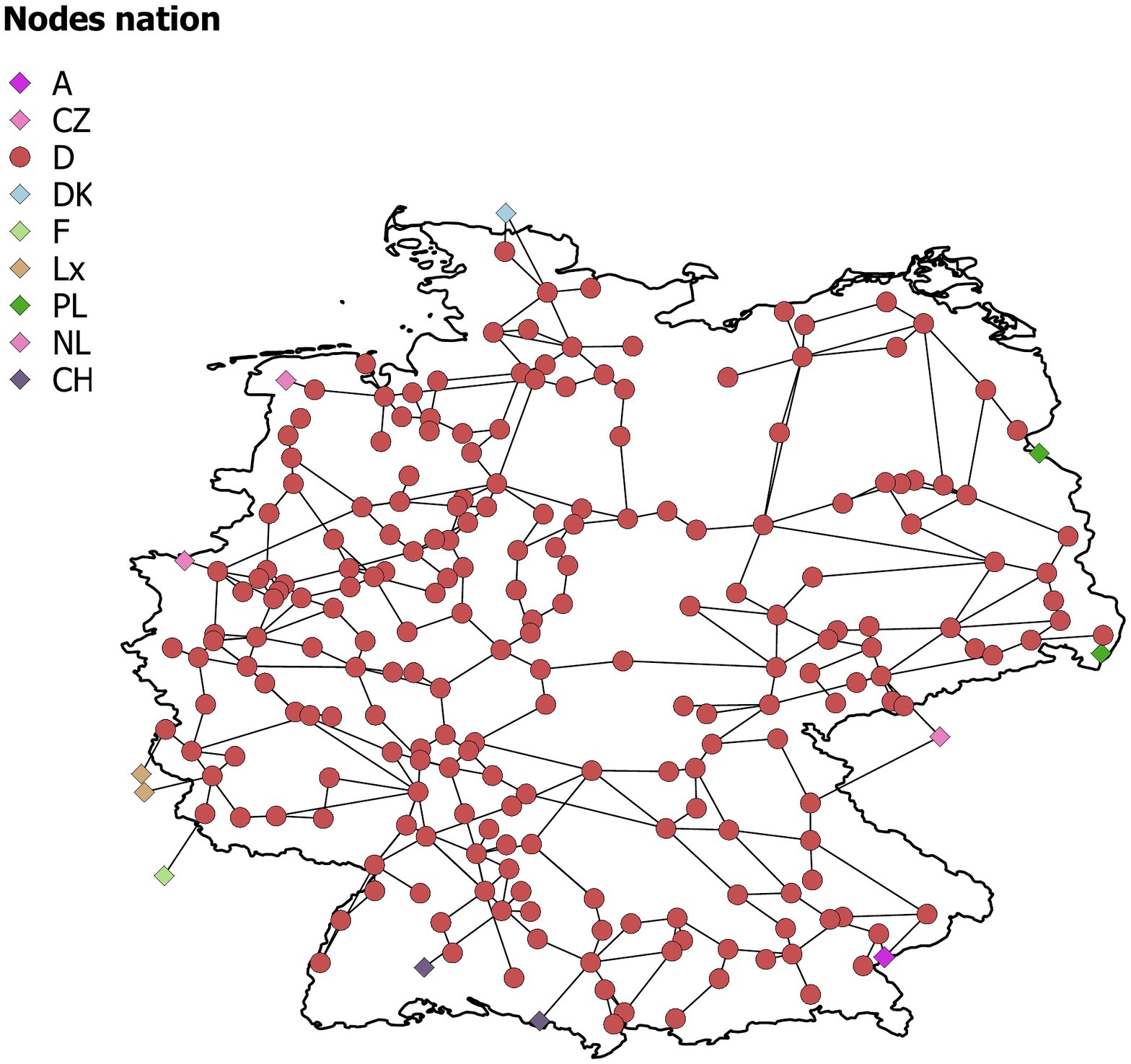} 
			\caption{A geographical representation of the German 380 KV grid. The German nodes are shown in red circles, the nodes connecting to foreign countries are represented by diamonds, coloured on the base of the country, which they connect to.} \label{fig:de_grid}
		\end{figure}

The proposed analysis represents an enrichment and extension of the dataset provided in \cite{Zhou:2005} and updated in \cite{Hutcheon2013}. This dataset contains a full DC working state of the European network (ex UCTE), in the presence of the real geographical distribution of RES in the system.
		The German Transmission Grid (GTG) dataset, whose geographical representation is shown in figure \ref{fig:de_grid}, is composed by 231 nodes and 304 edges. Also, information about 82 Conventional Generators (CGs) is given, together with the information regarding the amount of RES installed generation over each node. In the following, all the available information regarding all the network components is listed.
		\subsection*{Nodes}
		Each network node (also called bus) corresponds to a transmission substation, working at a nominal voltage of 380KV. For each node, it is available all the information needed for fully describing the system, extracted from the original dataset or by making use of the techniques described in the methods section. In particular:
		\begin{itemize}
			\item The geographical position was obtained by georeferentiation, by means of a geographical  reprojection of the original dataset. This methodology is able to produce a quite accurate positioning of the substations into the territory. In order to improve the precision of the positioning, if needed, a cross-matching with openstreetmap data is suggested. The resulting position is given in  lat/lon values.
			\item The nodes load (in MW),  was extracted from the original dataset in \cite{Zhou2005, Hutcheon2013}. The foreign connection nodes' loads were changed in order to take into account foreign power exchanges.
			
			\item Each node code and id are available and are the same as in the original dataset. This is useful for cross referencing between the different tables of the dataset.
			\item Each node conventional power output (in MW) is given, defined as the sum of all conventional generation on the node, referred to the actual dispatch.
			\item Each node installed wind and solar generation (in MW) are given. This information has been obtained by means of the analysis described in methods section.
			\item The information regarding the nodes rotating inertia (in $Kg\cdot m^2$), was estimated by means of the methodology described in the methods section.
		\end{itemize}
		\subsection*{Edges}
		For each of the 314 existing connections between the network  nodes, here called edges, the following information is given:
		\begin{itemize}
			\item The edge starting and ending nodes $i$ and $j$, given in terms of nodes ids and codes.
			\item The edge estimated length, obtained by calculating the geodetic distance between the edge terminal nodes. 
            \item The lines nominal powers $C_{ij}$.
			\item The edge reactance $X$, obtained from the original UCTE dataset. No resistance is given in this dataset, assuming negligible resistive processes on 380KV lines, but it can be estimated by assuming an average resistance per length and by multiplying this value with the given edge length.
		\end{itemize}
		\subsection*{Generators}
		Each of the 89 conventional generators present in the dataset is described by:
		\begin{itemize}
			\item id and code, same as the original UCTE dataset;
			\item id and code of the node at which the generator is attached;
			\item technology, obtained from the UCTE dataset;
			\item power output, obtained from the UCTE dataset;
			\item minimum and maximum nominal power, obtained from the UCTE dataset.
		\end{itemize}

The published dataset includes 5 directories, one for each value of  $P_\%$. Each directory contains 4 files, one containing the nodes data, called ''nodes.csv''; one containing the edges data, called ''branches.csv'', one containing the generation data, called ''generators.csv'' and the last one is the ''matpower.m'' file, which contains all the data needed for running a power flow analysis. It can be easily modified and loaded by matpower \cite{Zimmerman2011}.

% Tables should be used to support the data records, and should clearly indicate 
% the samples and subjects, their provenance, and the experimental manipulations 
% performed on each. They should also specify the data output resulting from each 
% data-collection or analytical step, should these form part of the archived record. 
% Please see the submission guidelines at the \emph{Scientific Data} website, and 
% our Word templates for more information on preparing such tables. 

\section*{Technical Validation}

% This section presents any experiments or analyses that are needed
% to support the technical quality of the dataset. This section may
% be supported by up figures and tables, as needed. This is a required
% section; authors must present information justifying the reliability
% of their data.

The presented data is obtained from two main databases:
\begin{itemize}
\item The German power grid data from \cite{Zhou:2005,Hutcheon2013};
\item The data about the distribution of RES generators, published in \cite{energymap};
\end{itemize}
The first dataset is part of a peer reviewed journal article\cite{Zhou:2005}, further enriched in \cite{Hutcheon2013}, presented as a conference proceeding. It is a common dataset used for power grid analysis, and contains the full representation of the UCTE (the old association of european TSOs, actually ENTSO-E \cite{ENTSO-E2012}) network. The second dataset is part of a project of a census of installed RES generation in Germany, supervised by the Bundesnetzagentur \cite{BNA}, the German federal agency for networks. 

The datasets have been processed by means of fully maintained programs, which are standard leading choices of the community. This has been done to ensure the maximum reliability of the performed analysis. In particular:
\begin{itemize}
\item For saving and processing the geographical data, I used PostGIS, the GIS extension for PostgreSQL databases;
\item Geographical analysis and visualization was performed with the suite QGIS \cite{QGIS};
\item The power flow analysis was performed with MATPOWER \cite{Zimmerman2011};
\item The data treatment, connection between different data process methods and I/O was done with python (Python Software Foundation, \url{https://www.python.org/}).
 
\end{itemize}

\section*{Usage Notes}

%Brief instructions that may help other researchers reuse these dataset.
%This is an optional section, but strongly encouraged when helpful
%to readers. This may include discussion of software packages that
%are suitable for analyzing the assay data files, suggested downstream
%processing steps (e.g. normalization, etc.), or tips for integrating
%or comparing this with other datasets. If needed, authors are encouraged
%to provide code, programs, or data processing workflows when they may help 
%others analyse the data. We encourage authors to archive related code in 
%a DOI-issuing archive when possible, but code may also be supplied as 
%supplementary information files. 

%VEDERE SE SCRIVERLO The proposed dataset can be easily used by accessing to the excel file. The data regarding loads consumption and generators production could be used for analysis includng power flow calculations such as matpower CITE, or internal 
The described dataset can be used for studying the RES impact on transmission systems from various points of view. In particular, it can be used for:
\begin{itemize}
\item Steady state analysis: The dataset contains all information needed for performing DC power flow analysis, DCOPF included. This makes this dataset very valuable to monitor the impact of RES on markets and congestions. The data contains the files ''grid.m'',  which can be read directly from MATPOWER \cite{Zimmerman2011}. It allows various types of studies, mainly based on the solution of power flow equations. Further information on how to read matpower files are given in the matpower manual \cite{Zimmerman2011}.
\item Dynamic analysis: the dataset contains all information needed for performing dynamical analysis based on the swing equation, Kuramoto models, and their stochastic counterparts. From this point of view, it can be a useful reference system for these analyses, useful for testing general results on a realistic network of large size.

\end{itemize}

\section*{Acknowledgements}
The author gratefully acknowledge the support from the Federal Ministry of Education and Research (BMBF grant no. 03SF0472D NET-538-167)). The author acknowledge the precious help and supervision of Prof. Meyer-Ortmanns. Any opinion, findings and conclusions or recommendations expressed in this material are those of the author and do not necessarily reflect the views of the funding parties.

% Text acknowledging non-author contributors. Acknowledgements should
% be brief, and should not include thanks to anonymous referees and
% editors, or effusive comments. Grant or contribution numbers may be
% acknowledged. Author contributions Please describe briefly the contributions
% of each author to this work on a separate line. 

% AK did this and that. 

% BG did this and that and the other. 

\section*{Competing financial interests}

The author declares  no competing financial interests.

\section*{Data Citations}
Mureddu, M. (2016). Representation of the German transmission grid for Renewable Energy Sources impact analysis.\emph{figshare}. http://doi.org/10.6084/m9.figshare.4233782.v2

% Bibliographic information for the data records described in the manuscript.

% 1. Lastname1, Initial1., Lastname2, Initial2., ...\& LastnameN, InitialN. \emph{Repository name} Dataset accession number or DOI (YYYY).

\end{document}